\documentclass[numberedappendix]{emulateapj}
\begin{document}

\slugcomment{ApJ accepted: March 18, 2008}
\title{Resolving the chemistry in the disk of TW~Hydrae \\
I. Deuterated species }
\author{Chunhua~Qi\altaffilmark{1}, David~J.~Wilner\altaffilmark{1},
Yuri~Aikawa\altaffilmark{2}, Geoffrey~A.~Blake\altaffilmark{3},
Michiel~R.~Hogerheijde\altaffilmark{4}}
\altaffiltext{1}{Harvard--Smithsonian Center for
Astrophysics, 60 Garden Street, MS 42, Cambridge, MA 02138,
USA; cqi@cfa.harvard.edu, dwilner@cfa.harvard.edu.}
\altaffiltext{2}{Department of Earth and Planetary Sciences, Kobe
University, Kobe 657-8501, Japan; aikawa@kobe-u.ac.jp.}
\altaffiltext{3}{Divisions of Geological \& Planetary Sciences and
Chemistry \& Chemical Engineering, California Institute of Technology,
MS 150--21, Pasadena, CA 91125, USA; gab@gps.caltech.edu.}
\altaffiltext{4}{Leiden Observatory, Leiden University, PO Box 9513, 2300
RA, Leiden, The Netherlands; michiel@strw.leidenuniv.nl.}
\begin{abstract}

We present Submillimeter Array (SMA) observations of several
deuterated species in the disk around the classical T Tauri
star TW~Hydrae at arcsecond scales, including detections of the DCN J=3--2
and DCO$^+$ J=3--2 lines, and upper limits to the HDO 3$_{1,2}$--2$_{2,1}$,
ortho-H$_2$D$^+$ 1$_{1,0}$--1$_{1,1}$ and para-D$_2$H$^+$
1$_{1,0}$--1$_{0,1}$ transitions.
We also present observations of the HCN J=3--2, HCO$^+$ J=3--2 and
H$^{13}$CO$^+$ J=4--3 lines for comparison with their deuterated
isotopologues.  We constrain the radial and vertical distributions
of various species in the disk by fitting the data using a model
where the molecular emission from an irradiated accretion disk
is sampled with a 2D Monte Carlo radiative transfer code.
We find that the distribution of DCO$^+$ differs markedly from that of
HCO$^+$. The D/H ratios inferred change by at least one order of magnitude
(0.01 to 0.1) for radii $<$30 AU to $\ge$70 AU and 
there is a rapid falloff of the abundance of DCO$^+$ at 
radii larger than 90 AU.
Using a simple analytical chemical model, we constrain the
degree of ionization, x(e-)=n(e-)/n(H$_2$), to be $\sim10^{-7}$
in the disk layer(s) where these molecules are present.
Provided the distribution of DCN follows that of HCN, the ratio of
DCN to HCN is determined to be 1.7$\pm$0.5 $\times$ 10$^{-2}$;
however, this ratio is very sensitive to the poorly constrained 
vertical distribution of HCN. The resolved
radial distribution of DCO$^+$ indicates that {\it in situ} deuterium
fractionation remains active within the TW~Hydrae disk and must be
considered in the molecular evolution of circumstellar accretion disks.

\end{abstract}


\section{Introduction}

Millimeter-wave interferometers have imaged the gas and dust
surrounding over a dozen T Tauri and Herbig Ae stars (see
\citealp{dutrey_g07} for a review). These studies demonstrate the
potential to dramatically improve our understanding of disk physical
and chemical structure, providing insights that will ultimately
enable a more comprehensive understanding of star and planet
formation. As analogs to the Solar Nebula, these circumstellar disks offer
a unique opportunity to study the conditions during the planet formation
process, especially the complex chemical evolution that must occur.
In the outer parts of the disk directly accessible to millimeter-wave
interferometry, observations of deuterated species are particularly
important because they can constrain the origin of primitive solar system
bodies such as comets and other icy planetesimals.

Deuterated molecule chemistry is sensitive to the temperature history
of interstellar and circumstellar gas, as well as to density-sensitive
processes such as molecular depletion in the cold  ($<$ 20 K) and
dense ($>$ 10$^6$ cm$^{-3}$) disk midplane. The similarity of
molecular D/H ratios between comets and high mass hot cores has
been used to argue for an ``interstellar'' origin of cometary matter, but
ambiguity remains and this argument is not secure
(\citealp{bergin_a07}). \citet{aikawa_h99} investigate the chemistry
of deuterium-bearing molecules in outer regions of protoplanetary
disks and find that molecules formed in the disk have similar D/H
ratios as those in comets.  
To date, the determination of D/H ratios in disks has been limited to the 
measurements of DCO$^+$/HCO$^+$ in two classical T Tauri Stars (cTTs) with 
single dish telescopes (TW~Hydrae, van Dishoeck et al. 2003; DM~Tauri, 
Guilloteau et al. 2006), but DCO$^+$ has not been observed in comets. 
Although H$_2$D$^+$ and HDO lines have been detected in disks 
(Ceccarelli et al. 2004, 2005), there are no corresponding H$_3^+$ 
and H$_2$O observations that allow derivations of the D/H ratio.
Therefore, direct measurements in disks of species 
observed in comets, such as DCN and HCN, will be important 
to help relate deuterium fractionation in disks to that in comets.

Deuterated molecules in cold pre-stellar and protostellar cores 
are found to be enhanced by orders of magnitude over the elemental 
D/H abundance ratio of 1.5$\times$10$^{-5}$ through fractionation 
in the gas phase at low temperatures, an effect driven primarily 
by deuterated H$_3^+$, which is formed by exchange reaction between 
H$_3^+$ and HD, then transfers its deuteron to neutral species.
Theoretical models of disks (\citealp{aikawa_v02,willacy07}) with
realistic temperature and density structure show that the DCO$^+$/HCO$^+$
ratio increases with radius due to the decreasing temperature moving
out from the star.  Spatially resolved data of deuterated molecules will
help test disk physical models through these chemical consequences.

The current capabilities of millimeter-wave observatories are limited
both by sensitivity and by the small angular size of circumstellar
disks, which makes spatially resolving the deuterium fractionation difficult.
In this work we take advantage of the proximity of the most nearby  
classical T Tauri star, TW~Hydrae, which is surrounded by a disk of 
radius $3\farcs5$, or 200 AU at a distance of 56 pc (\citealp{qi_h04}), 
and a possible orbiting planet of mass (9.8$\pm$3.3) M$_{Jupiter}$ 
at 0.04 AU (\citealp{setiawan_h08}), to study the physical 
and chemical structure of a protoplanetary environment at high spatial 
and spectral resolution using interferometry. The TW~Hydrae disk is viewed
nearly face-on, and so is well posed to investigate the radial
distribution of molecular emission. Although the current
angular resolution of $\sim$1--2$''$ places only a few independent
beams ( or `pixels') across the disk, effectively improved resolution
of the disk chemistry can be achieved thanks to the fact that the gas
kinematics are essentially Keplerian.~\citet{beckwith_s93} show that 
molecular line emission from Keplerian disks displays a ``dipole'' pattern 
near the systematic velocity due to the shear created by the orbital motion.
The emission near the systemic velocity is dominated by 
the outer disk regions, and the separation of the emission peaks depends
sensitively on the abundance gradient of the emitting species with radius.
This important feature, commonly seen in velocity channel maps of disks 
with a range of inclination angles (even as small as 6--7 degrees in the 
case of TW~Hydrae, \citealp{qi_h04}),
may be used to study the radial distribution of molecular emission far more
precisely than is afforded by the limited number of ``pixels'' provided by
the available resolution.

Here we report on Submillimeter Array
(SMA)\footnote{The Submillimeter Array is a joint
project between the Smithsonian Astrophysical Observatory and the
Academia Sinica Institute of Astronomy and Astrophysics, and is
funded by the Smithsonian Institution and the Academia Sinica.}
(\citealp{ho_m04})
observations of deuterated species in the disk around TW~Hydrae, including
the first detection and images of DCN and DCO$^+$.
In \S 2 we describe the observations,
while in \S 3 we introduce the analysis method used and the molecular
distribution parameters derived. We describe the model fitting results and 
discuss the implications in \S 4, and we present a summary and conclusions in \S 5.

\section{Observations}

All of the observations of TW~Hydrae
(R.A.: 11$^h$01$^m$51.$^s$875; Dec: -34$^\circ$42$'$17.$''$155; J2000.0)
were made between 2005 February and 2006 December using the SMA 8 antenna 
interferometer located atop Mauna Kea, Hawaii.  Table 1 and Table 2 summarize the
observational parameters for the detected and undetected species, respectively.
The SMA receivers operate in a double-sideband (DSB) mode with an
intermediate frequency band of 4--6 GHz which is sent over fiber optic
transmission lines to 24 ``overlapping'' digital correlator ``chunks''
covering a 2 GHz spectral window. Two settings were used for observing
the HCN/DCN and HCO$^+$/DCO$^+$ lines: at 265.7--267.7 GHz
(lower sideband) the tuning was centered on the HCN J=3--2 line at
265.8862 GHz in chunk S22 while the HCO$^+$ J=3--2 line at 267.5576 GHz was
simultaneously observed in chunk S02; at 215.4--217.4 GHz (lower sideband)
the tuning was centered on the DCO$^+$ J=3--2 line at 216.1126 GHz in
chunk S16 while the DCN J=3--2 line at 217.2386 GHz was
simultaneously observed in chunk S02 (the HDO 3$_{1,2}$--2$_{2,1}$ line
at 225.90 GHz was also covered in the USB in chunk S06).
Combinations of two array configurations (compact and extended) were
used to obtain projected baselines ranging from 6 to 180 meters.
Cryogenic SIS receivers on each antenna produced system temperatures
(DSB) of 200-1400 K at 260 GHz and 100--200 K at 210 GHz.
Observations of ortho-H$_2$D$^+$ and H$^{13}$CO$^+$ were simultaneously
carried out (in the dual receiver observational mode) on 28 December 2006 using only
the compact array configuration. Three out of eight antennas were equipped
with 400 GHz receivers (DSB system temperature between 400 and 800 K)
and were available for observations of the 372.4213 GHz
ortho-H$_2$D$^+$~1$_{1,0}$--1$_{1,1}$ line, and all eight antennas
were operating with 345 GHz receivers (DSB system temperature
between 150 and 300 K) for observations of the
346.9985 GHz H$^{13}$CO$^+$ J=4--3 and 345.796 GHz CO J=3--2 lines.
The correlator was configured with a narrow band of 512 channels over
the 104~MHz chunk, which provided 0.2 MHz frequency resolution, or
0.28 km~s$^{-1}$ at 217 GHz, 0.23 km~s$^{-1}$ at 267 GHz and
0.16 km~s$^{-1}$ at 372 GHz.
Observations of the 691.6604 GHz para-D$_2$H$^+$~1$_{1,0}$--1$_{0,1}$
line were shared with observations of the CO J=6--5 line
made on 17 February 2005 (details provided in \citealp{qi_w06}).
Calibrations of the visibility phases and amplitudes were achieved with
interleaved observations of the quasars J1037-295 and J1147-382, typically
at intervals of 20-30 minutes. Observations of Callisto provided the
absolute scale for the calibration of flux densities. The uncertainties in
the flux scale are estimated to be 15\%. All of the calibrations were
done using the MIR software package
\footnote{http://www.cfa.harvard.edu/$\sim$cqi/mircook.html}, while
continuum and spectral line images were generated and CLEANed
using MIRIAD.

\begin{table*}
\centering
\caption{Observational Parameters for SMA TW~Hydrae (detected species)}
\begin{tabular}{lccccc}
\hline
&HCN 3--2&HCO$^+$ 3--2&H$^{13}$CO$^+$ 4-
-3 &DCN 3--2$^c$&DCO$^+$ 3--2 \\
\hline
\hline
Rest Frequency (GHz):& 265.886 & 267.558& 346.999 & 217.239& 216.113\\
Observations   & 2005 Mar 04 & 2005 Mar 04 & 2006 Dec 28 & 2006 Apr 28
& 2006 Apr 28 \\
               & 2005 Apr 21 & 2005 Apr 21 & &            & 2006 Feb 03 \\
               & 2005 April 26 & 2005 Apr 26 & &            & \\
Antennas used  & 7 & 7 & 8 & 8 & 8 \\
Synthesized beam: & $1\farcs6 \times 1\farcs1$ PA -0.5$^{\circ}$
                  & $1\farcs6 \times 1\farcs1$ PA -6.3$^{\circ}$
          & $4\farcs1 \times 1\farcs8$ PA  3.3$^{\circ}$
                  & $5\farcs9 \times 3\farcs2$ PA -1.5$^{\circ}$
                  & $2\farcs6 \times 1\farcs6$ PA  2.8$^{\circ}$ \\
Channel spacing (km s$^{-1}$): & 0.23 & 0.23 km & 0.70 & 0.56 & 0.28 \\
R.M.S.$^a$ (Jy beam$^{-1}$):   & 0.35 & 0.29 &
0.16 & 0.10 & 0.10 \\
Peak intensity$^b$ (Jy) & 4.7 & 2.0 & 0.70 & 0.31& 0.56\\
\hline
\end{tabular}
\tablenotetext{a} { SNR limited by the dynamic range. }
\tablenotetext{b} { Intensity averaged over the corresponding beam.}
\tablenotetext{c} { Only compact configuration data used for DCN observation.}
\end{table*}
 
\begin{table*}
\centering
\caption{Observational Parameters for SMA TW~Hydrae (undetected species)}
\begin{tabular}{lccc}
\hline
&HDO 3$_{1,2}$--2$_{2,1}$&o-H$_2$D$^+$
1$_{1,0}$--1$_{1,1}$&p-D$_2$H$^+$ 1$_{1,0}$--1$_{0,1}$\\
\hline
\hline
Rest Frequency (GHz):& 225.897 & 372.421 & 346.999 \\
Observations:   & 2006 April 28 & 2006 Dec 28 & 2005 Feb 17\\
Antenna used:   &   8  &  3  &  4 \\
Synthesized beam: & $5\farcs7 \times 3\farcs1$ PA -0.6$^{\circ}$
                  & $4\farcs7 \times 3\farcs8$ PA 14.5$^{\circ}$
          & $3\farcs3 \times 1\farcs3$ PA 7.5$^{\circ}$\\
Channel spacing (km s$^{-1}$): & 0.54 & 0.65 & 0.35 \\
R.M.S. (Jy beam$^{-1}$):   & 0.11 & 1.17  & 7.39  \\
N$_{molecule}$ (1 $\sigma$) (cm$^{-2}$): & $<$2.0 $\times$ 10$^{14}$
& $<$1.7 $\times$ 10$^{12}$  & $<$9.0 $\times$ 10$^{14}$ \\
\hline
\end{tabular}
\end{table*}

\section{Spectral Line Modeling}

Several model approaches have been developed to interpret interferometric
molecular line and continuum observations from disks (\citealp{dutrey_g07}).
In brief, the approach of Qi et al. (2004, 2006) works as follows: the 
kinetic temperature and density structure of the disk is determined by modeling
the spectral energy distribution (SED) assuming well mixed gas and dust,
with the results confirmed by the resolved (sub)mm continuum images. Then,
a grid of models with a range of disk parameters including the outer radius 
R$_{out}$, the disk inclination $i$,
position angle P.A. and the turbulent line width v$_{turb}$ are produced and a 2D accelerated Monte Carlo model
(\citealp{hogerheijde_v00}) is used to calculate the radiative transfer and
molecular excitation. 
The collisional rates are taken from the Leiden Atomic and Molecular Database 
\footnote{http://www.strw.leidenuniv.nl/$\sim$moldata} (Schoier et al. 2005) 
for non-LTE line radiative transfer calculations. Specifically in 
the models used in this paper, 
the rate coefficients for HCO$^+$ and DCO$^+$ in collisions with H$_2$
have been calculated by Flower (1999); the rate coefficients for HCN
and DCN in collisions with H$_2$ have  
been scaled from the rate of HCN-He calculated by Green \& Thaddeus (1974). 
The model parameters are fitted 
using a $\chi^2$ analysis in the $(u,v)$ plane.

In the studies by \citet{qi_h04,qi_w06}, the distribution of CO molecules 
was assumed to
follow the H nuclei as derived from the dust density structure and a gas of
solar composition.  However, molecules in disks do not
necessarily share the same distribution as molecular hydrogen. 
Theoretical models
(e.g. \citealp{aikawa_m96,aikawa_n06}) predict so-called three-layered
structure; most molecules are
photo-dissociated in the surface layer of the disk and frozen out in the
mid-plane where most of hydrogen resides, with an abundance that peaks in
the warm molecular layer at intermediate scale heights.
To approximate this more complex behavior, we introduce new molecular 
distribution parameters for use in spectral line modeling.

For first-order analysis in the radial molecular distributions, the column 
densities are assumed to vary as a power law as a function of radius:
\begin{eqnarray}
\nonumber
\Sigma_i (r)=\Sigma_i (10{\rm AU})(\frac{r}{10{\rm AU}})^{p_i}
\end{eqnarray}
Here $\Sigma_i$ and $p_i$ describe column density distribution of a
specific species ($i$) rather than disk (hydrogen) column density.
In most disk models to date, a single radial power-law is adopted for 
fitting the hydrogen or dust surface density. By assuming CO follows the 
distribution of hydrogen column density, \citet{qi_h04,qi_w06} also find
satisfactory power-law fits for multiple CO transtions.
When a single power law is insufficient for fitting the 
radial distribution, then a broken power law with two different 
indices will be used.

\begin{figure}[htb]
\centering
\includegraphics[width=0.45\textwidth]{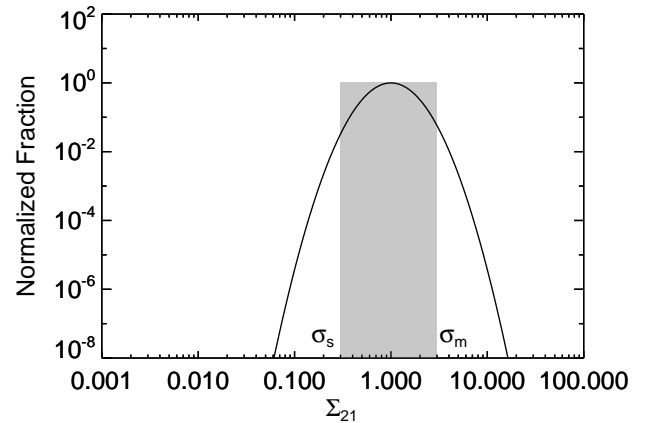}
\caption{ This schematic diagram shows an arbitrary molecular vertical
distribution as  a function of $\Sigma_{21}$ measured from the disk
surface at a certain radius. $\sigma_m$ and $\sigma_s$ are the
midplane and surface boundary parameters for model fitting (see \S3 of
the text).
 \label{fig:vertdiag}}
\end{figure}

\begin{figure*}
\centering
\includegraphics[width=0.55\textwidth]{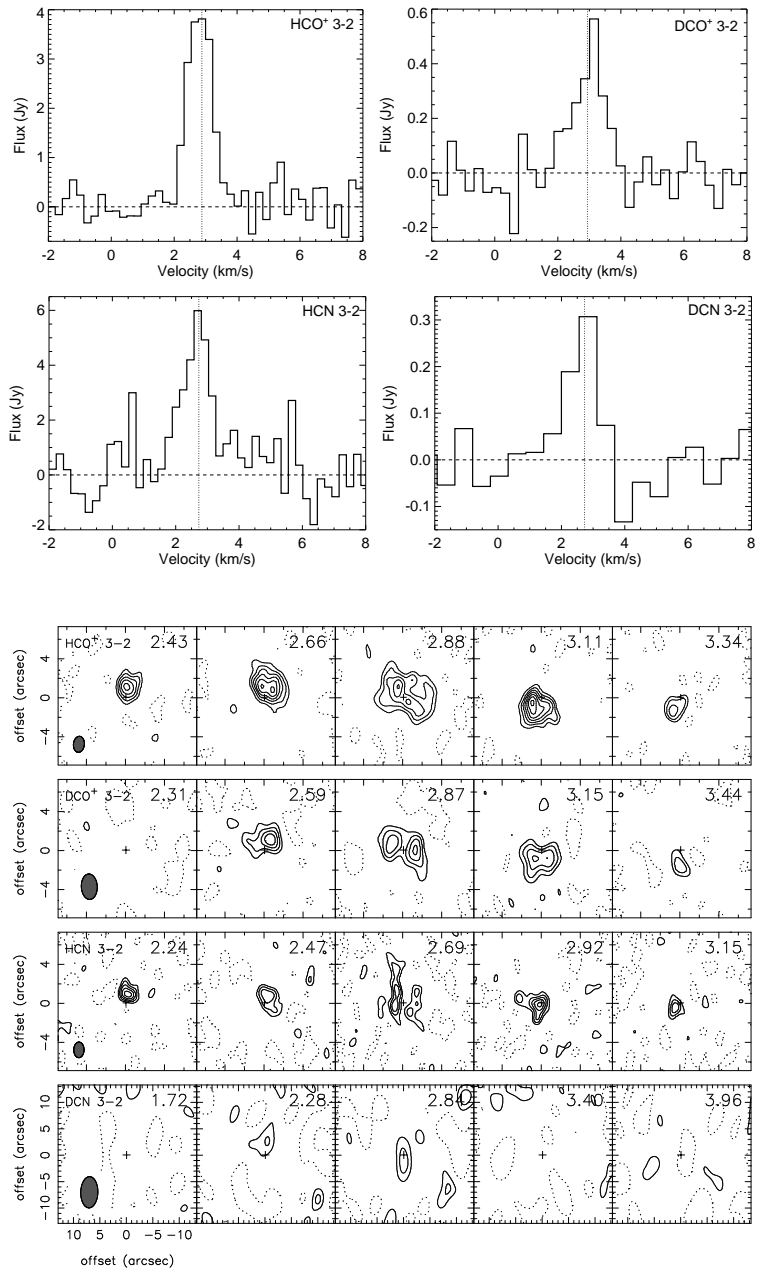}
\caption{ Top: The HCO$^+$, DCO$^+$, HCN and DCN J=3--2 spectra at the
peak continuum (stellar) position. The fluxes of HCO$^+$
and DCO$^+$ are averaged over the beam of DCO$^+$ J=3--2
($2\farcs6 \times 1\farcs6$ PA  2.8$^{\circ}$).
The fluxes of HCN and DCN are averaged over the beam of
DCN J=3--2 ($5\farcs9 \times 3\farcs2$ PA -1.5$^{\circ}$).
The vertical dotted lines indicate the positions of the fitted V$_{LSR}$ for
each molecular transition except for DCN 3-2 where we adopt the V$_{LSR}$
from that of HCN J=3--2.
Bottom: Velocity channel maps of the HCO$^+$, DCO$^+$, HCN and DCN
J=3--2 emissions toward TW Hydrae. The angular resolutions are
$1\farcs6 \times 1\farcs1$ at PA -6.3$^{\circ}$ for HCO$^+$ J=3--2 and
$1\farcs6 \times 1\farcs1$ at PA -0.5$^{\circ}$ for HCN J=3--2.
The cross indicates the continuum
(stellar) position. The axes are offsets from the pointing center in
arcseconds. The 1$\sigma$ contour steps are 0.4, 0.12, 0.35 and 0.09
Jy~beam$^{-1}$ for HCO$^+$, DCO$^+$, HCN and DCN J=3--2 respectively
and the contours start at 2$\sigma$.
 \label{fig:specdata}}
\end{figure*}

To calculate the surface density at different radii, the vertical
molecular distribution is needed, but this distribution may well
vary with distance from the star. However, theoretical models
(\citealp{aikawa_n06})
indicate that the vertical distribution of molecules at different
radii is similar as a function of the hydrogen column density measured
from the disk surface,
$\Sigma _{21}\equiv \Sigma _{\rm H}/(1.59\times 10^{21}~{\rm cm}^{-2})$,
where the denominator is the conversion factor of the hydrogen column
density to $A_{v}$ for the case of interstellar dust. As indicated
by Figure 5 of \citet{aikawa_n06}, the vertical distribution of
molecular abundances shows a good correlation with $\Sigma _{21}$.
Figure~\ref{fig:vertdiag} shows a schematic diagram of an arbitrary
distribution of a molecule in disk where the x-axis shows $\Sigma _{21}$ and
the y-axis shows the normalized molecular fraction.
Smaller $\Sigma _{21}$ values (to the left of the plot) approach the surface of
the disk, while larger $\Sigma _{21}$ values (to the right) approach
the midplane of the disk. For modeling, we make the further simplifying
assumption that gaseous molecules exist with a constant abundance in layers
between $\sigma _s$ and $\sigma_m$, the surface and midplane boundaries of
$\Sigma _{21}$, respectively, as indicated in Figure~\ref{fig:vertdiag}
by the shaded area. 
While the vertical distribution of molecules in disks may have a more
complex distribution than assumed here, the adopted parameters provide a 
gross approximation to the vertical location where the species is most 
abundant. This is adequate for a first description, given the quality of
the data available.  For example, the model of Aikawa \& Nomura (2006) 
shows two vertical peaks of HCO$^+$, but the secondary peak is an order of 
magnitude smaller, and provides effectively negligible emission.
Also, at least for the nearly face-on disk of TW~Hydrae, the uncertainties 
in vertical distributions do not affect the derived radial distribution.

This simple model captures the basic characteristics of 
three-layered structure predicted by theoretical models. 
Using the new distribution parameters: $\Sigma_i (10{\rm AU})$ and
$p_i$ (radial), $\sigma _s$ and $\sigma_m$ (vertical), the radial and
vertical distributions of the molecules in disks can be constrained
by observation. 
The best fit model is obtained by minimizing the $\chi^2$, the weighted 
difference between the real and imaginary part of the complex visibility 
measured at the selected points of the ($u,v$) plane. The $\chi^2$ values 
are computed by the simultaneous fitting of channels covering LSR velocities 
from 1 to 4 km s$^{-1}$.  R$_{out}$ and $p_i$ are calculated on grids 
in steps of 5 (AU) and 0.2, respectively. 
$\log \left( \sigma_{s}\right)$ and  
$\log \left( \sigma _{m}\right)$ are calculated on grids in steps of 0.2 
within the range from $-$2 to 2. $\Sigma_i (10{\rm AU})$ and $i$ are
found with each pair of  
radial and vertical distribution parameters by minimizing $\chi^2$. 
The systemic velocity V$_{LSR}$ is fit separately since it is not correlated 
with those distribution parameters. For each fit parameter, 
the 1$\sigma$ uncertainties are estimated as $\chi ^{2}_{1\sigma }=\chi
^{2}_{m}+\sqrt{2n}$, where $n$ is the number of degrees of freedom and 
$\chi^2_m$ is the $\chi^2$ value of best-fit model, as in \citet{isella_t07}.
We tested values for the turbulent velocity in the range from 0.0 to 
0.15 km s$^{-1}$ and found that exact value does not have a significant 
impact, in part because of the coarse spectral resolution of the data.
Therefore, we fixed the turbulent velocity at an intermediate value, 
0.08 km s$^{-1}$. 

\section{Results and Discussion}

\begin{figure}[htb]
\centering
\includegraphics[width=0.35\textwidth,angle=90]{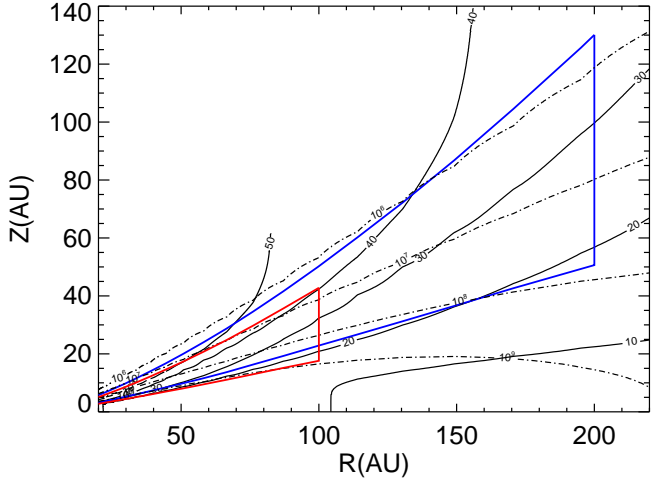}
\caption{ The solid and dotted contours show the temperature and
density profiles from the TW Hya model of \citet{calvet_d02}. The blue
and red lines confine the locations of HCO$^+$ and HCN in the model
(see text).
 \label{fig:modelvert}}
\end{figure}

\begin{figure*}
\centering
\includegraphics[width=0.55\textwidth,angle=90]{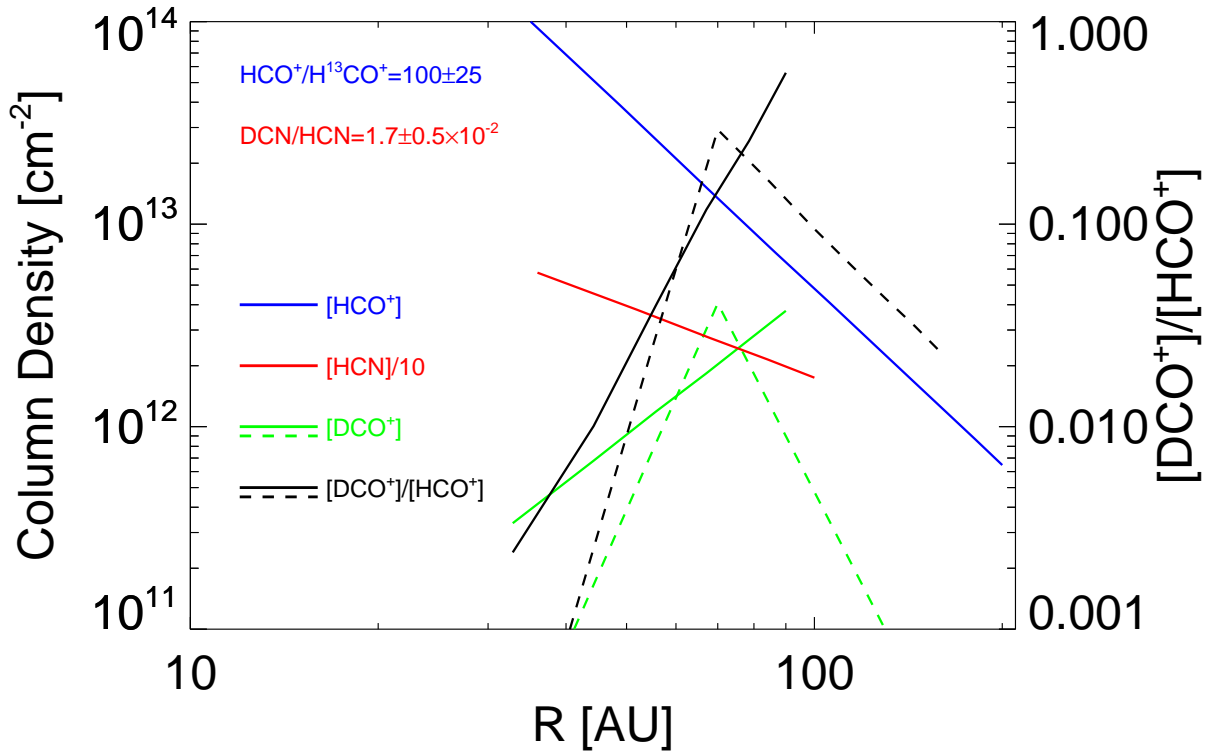}
\caption{ Radial distribution of molecular column densities
and DCO$^+$/HCO$^+$ ratio for the best-fit models for TW~Hydrae.
Solid lines depict single power-law fits, while
the dashed lines are for DCO$^+$ Model 3 where two power-law
indices are used.
 \label{fig:coldens}}
\end{figure*}


Figure~\ref{fig:specdata} shows the spectra of HCO$^+$, DCO$^+$, HCN
and DCN at the stellar (peak continuum) position and their velocity
channel maps.  
Table~3 summarizes the power-law fitting results on HCO$^+$,
DCO$^+$ and HCN.  
The DCN and H$^{13}$CO$^+$ lines are too weak for a $\chi^2$ analysis 
of their distribution parameters, and so only the ratios of DCN/HCN and
HCO$^+$/H$^{13}$CO$^+$ are fit.
Figure~\ref{fig:modelvert} shows the density and temperature contours
of the disk model (adopted from \citealp{calvet_d02}) and the locations of 
HCO$^+$ and HCN derived from the model fitting procedure.
Figure~\ref{fig:coldens} shows the radial distribution of the molecular
column densities of the best fit models for HCO$^+$, DCO$^+$ and HCN.
Not surprisingly, we found that the radial distributions are
better constrained than the vertical ones: the local minimum of $\chi^2$
for different grids of vertical parameters are within the noise limit,
i.e. the vertical parameters are the least well determined, due in part
to the face-on nature of the TW~Hydrae disk. We therefore treat the best-fit
vertical results as fixed, and investigate the uncertainty of other
parameters.

\begin{table*}
\caption{Fitting results}
\begin{tabular}{lcccc}
\hline
Parameters &HCO$^+$
&DCO$^+$$^a$ &DCO$^+$$^b$ &
HCN \\
\hline
\hline
Stellar Mass: M$_*$(M$_\odot$)& 0.6 & 0.6 & 0.6 & 0.6\\
Inclination: $i$(deg) & 6.8$\pm$0.3 & 7.4$\pm$0.6 & 7.4 & 6.6$\pm$0.8 \\
Systemic velocity: V$_{LSR}$(km s$^{-1}$) & 2.88$\pm$0.05 & 2.94$\pm$0.06 & 2.9
4 &
2.73$\pm$0.06 \\
Position angle: P.A.(deg) & -27.4 & -27.4 & -27.4 & -27.4\\
Turbulent line width: v$_{turb}$(km s$^{-1}$) & 0.08 & 0.08 & 0.08 & 0.08 \\
Outer radius: R$_{out}$(AU) & 200 $\pm$ 10 & 90 $\pm$ 5 & 160 & 100 $\pm$ 10\\
Column Density at 10AU: $\Sigma(10{\rm AU})$ (cm$^{-2}$) & 3.8 $\pm$0.5 $\times
$ 10$^{15}$ &
1.9$\pm$0.2 $\times$ 10$^{10}$ & 4.8 $\times$ 10$^6$ & 2.4$\pm$0.4
$\times$ 10$^{14}$ \\
Radial power index: p & -2.9 $\pm$ 0.3 &  2.4 $\pm$ 0.8 & 7,-6 $\tablenotemark{
c}$ & -1.0 $\pm$ 1.2 \\
Vertical Parameters: $\sigma_s$,$\sigma_m$ & 0.1, 10 & 0.1, 10 &
0.1,10 & 0.3, 30 \\
Minimum $\chi^2$: & 329588 & 575347 & 575347 & 297198 \\
Reduced $\chi^2$: $\chi^2_r$ & 1.21  & 1.95  & 1.95 & 1.35 \\
\hline
\end{tabular}
\tablenotetext{a} {DCO$^+$ Model 2.}
\tablenotetext{b} {DCO$^+$ Model 3, no error estimation.}
\tablenotetext{c} {DCO$^+$ turning point at radius 70 AU.}
\end{table*}

\subsection{HCO$^+$ and DCO$^+$}

The first detection of DCO$^+$ in the disk around TW~Hydrae
was obtained by van Dishoeck et al. (2003) with the James Clerk Maxwell
Telescope (JCMT), who reported a beam-averaged DCO$^+$/HCO$^+$
abundance ratio of 0.035. Our spatially resolved observations of 
the DCO$^+$ and HCO$^+$ J=3--2 emission from the disk suggest a 
more complex chemical situation.

\begin{figure*}
\centering
\includegraphics[width=0.9\textwidth]{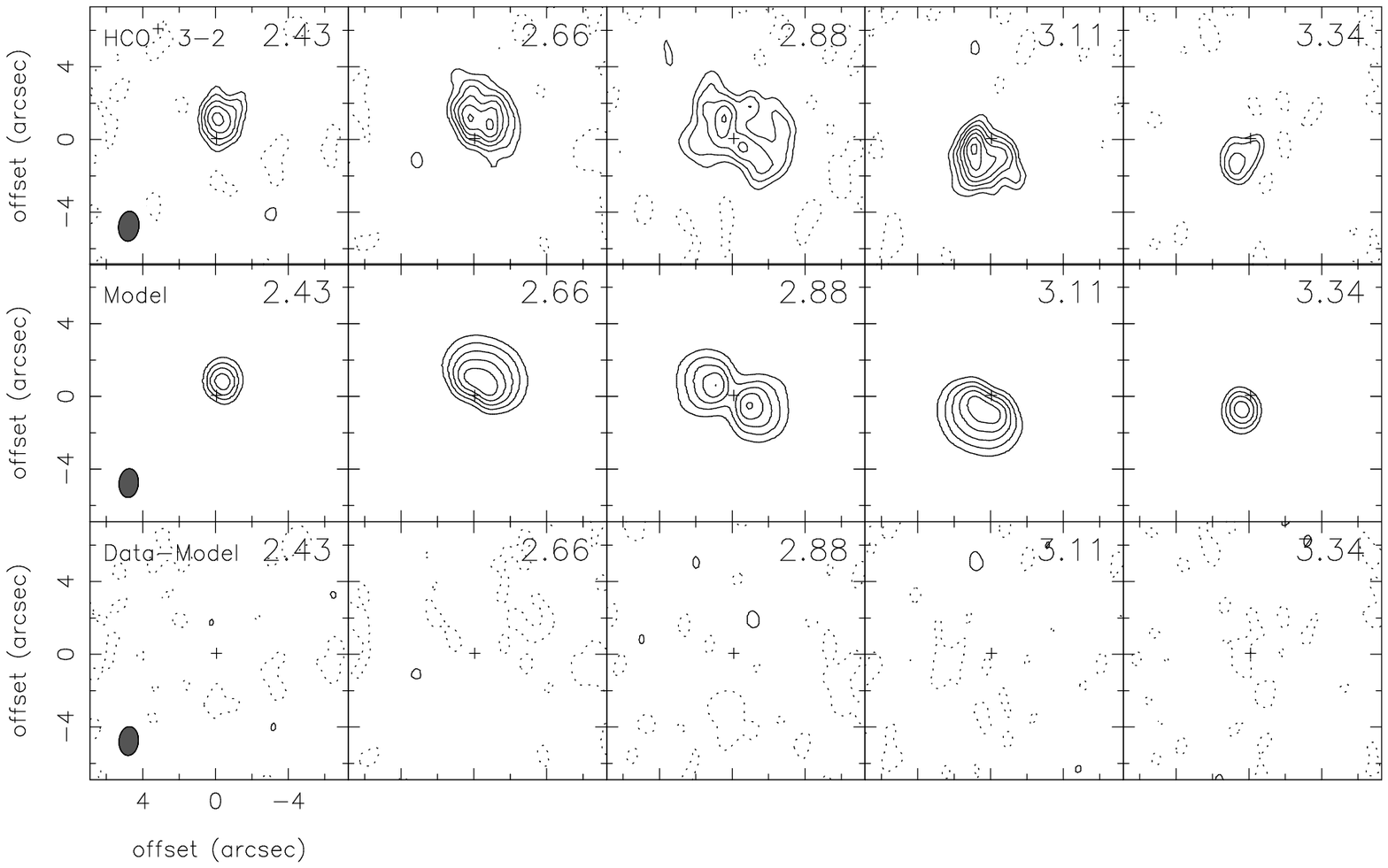}
\caption{Top: Velocity channel maps of the HCO$^+$
J=3--2 emission toward TW~Hydrae. The angular resolution is
$1\farcs6 \times 1\farcs1$ at PA -6.3$^{\circ}$.
The cross indicates the continuum (stellar) position.
The axes are offsets from the pointing center in arcseconds.
The 1$\sigma$ contour step is 0.4 Jy~beam$^{-1}$ and the
contours start at 2$\sigma$. Middle: channel map of the
best-fit model with the same contour levels. Bottom: difference
between the best-fit model and data on the same contour scale.
 \label{fig:hcomap}}
\end{figure*}

\begin{figure}[htb]
\centering
\includegraphics[width=0.45\textwidth]{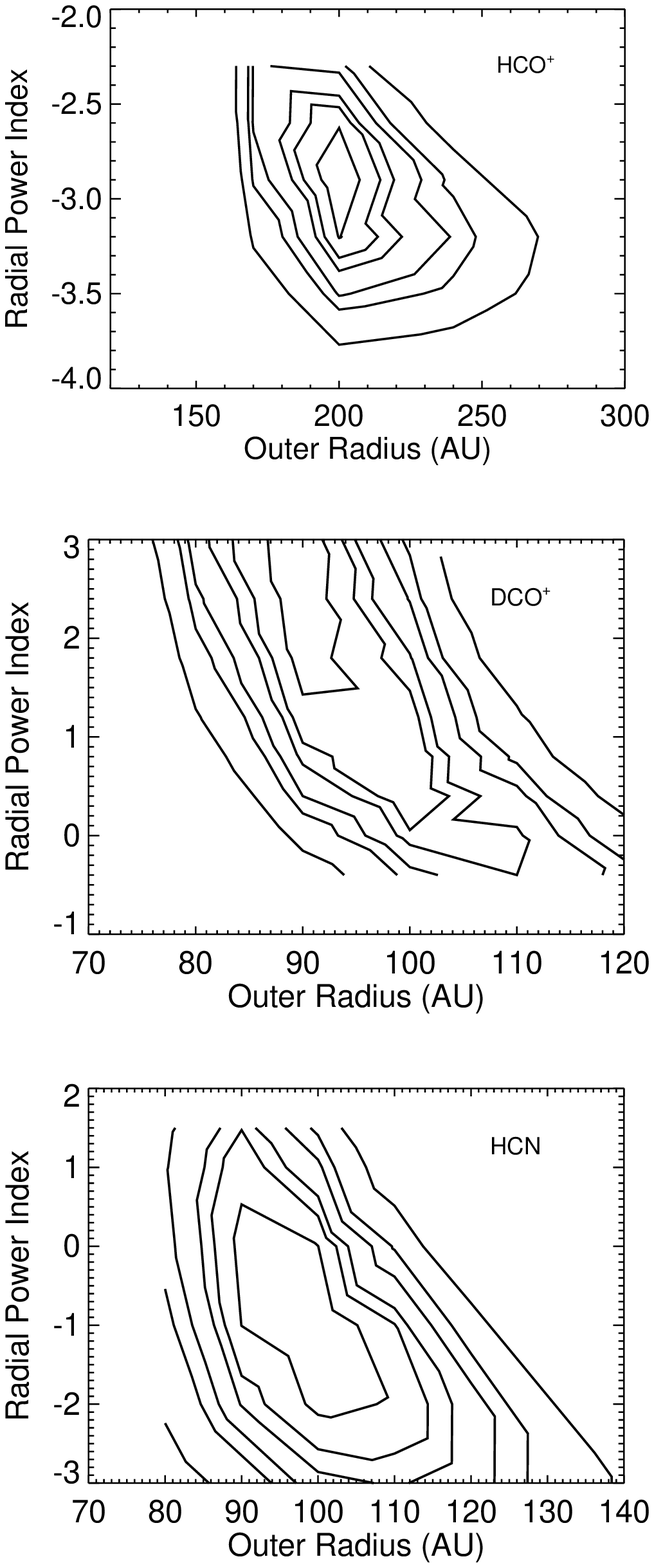}
\caption{ Iso-$\chi^2$ surfaces of (R$_{out}$,p$_i$)
for HCO$^+$, DCO$^+$ (as in Model 2) and HCN. Contours correspond to
the 1 to 6 $\sigma$ errors. For DCO$^+$, the index values
larger than 3 at around 90 AU indicate the ratios of
DCO$^+$/HCO$^+$ larger than 1, so the $\chi^2$ surface is not
calculated beyond that.
 \label{fig:chi}}
\end{figure}

\begin{figure}[htb]
\centering
\includegraphics[width=0.35\textwidth,angle=90]{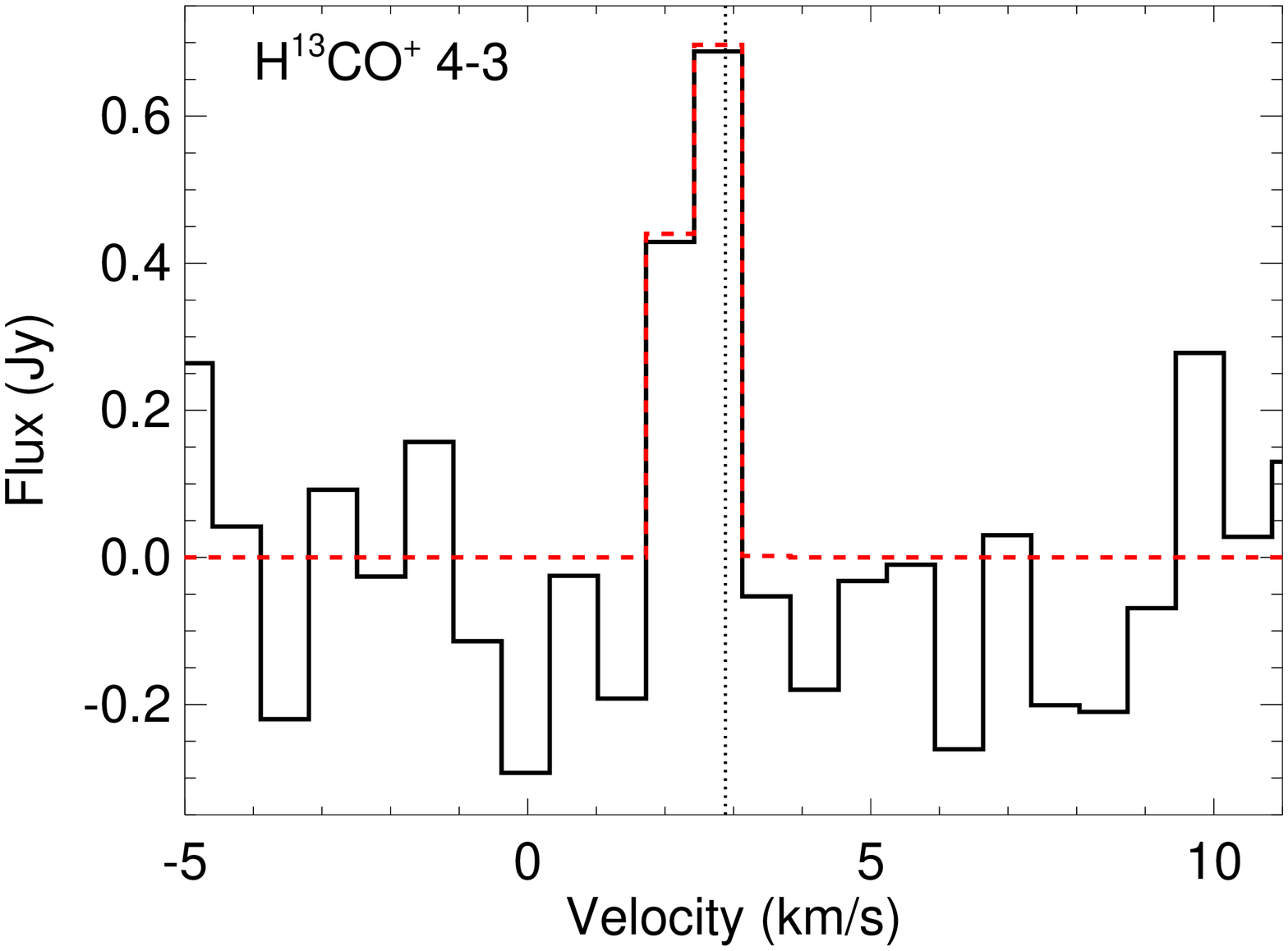}
\caption{ The beam-averaged H$^{13}$CO$^+$ J=4--3 spectra at the continuum (ste
llar)
position. The SMA data are presented by the solid histogram, and
the simulated model by the dashed histogram. The vertical dotted line
indicates the position of the fitted V$_{LSR}$ for HCO$^+$ J=3--2.
 \label{fig:h13co43}}
\end{figure}

Figure~\ref{fig:hcomap} shows the channel maps of HCO$^+$ J=3--2,
together with the best-fit model and the data-model residuals.
Table~3 lists the best-fit model parameters.
The $\chi^2$ surface for the radial power index p$_{\rm HCO^+}$ and the
outer radius R$_{out}$ is shown in the top panel of
Figure~\ref{fig:chi}. The 1$\sigma$ contour confines R$_{out}$ to 
200$\pm$10 AU and p$_{\rm HCO^+}$ to $-$2.9$\pm$0.3. 
The H$^{13}$CO$^+$ J=4--3 line was also detected, but the emission is
not strong enough to constrain its distribution.
Assuming that H$^{13}$CO$^+$ has the same distribution as HCO$^+$,
then fitting the ratio of HCO$^+$/H$^{13}$CO$^+$ to match the intensity of
the H$^{13}$CO$^+$ emission indicates an HCO$^+$/H$^{13}$CO$^+$ ratio of
100$\pm$25. This value is consistent with the nominal solar system 
value of 89.  
Figure~\ref{fig:h13co43} presents the best-fit model spectra of
H$^{13}$CO$^+$ J=4--3 overlayed on the SMA data.

The vertical distribution of DCO$^+$ is even less well
constrained than that of HCO$^+$, and we choose to adopt the same values 
of $\sigma_s$ and $\sigma_m$ for both HCO$^+$ and DCO$^+$ (as
indicated in Table~3). Because of the nearly face-on viewing
geometry for the disk, the fitting of radial distribution power-law
index p$_i$ is not affected significantly by the uncertainties 
in vertical structure, but only the value of $\Sigma_i$(10AU) 
needs to be adjusted. 

\begin{figure}[htb]
\centering
\includegraphics[width=0.35\textwidth,angle=90]{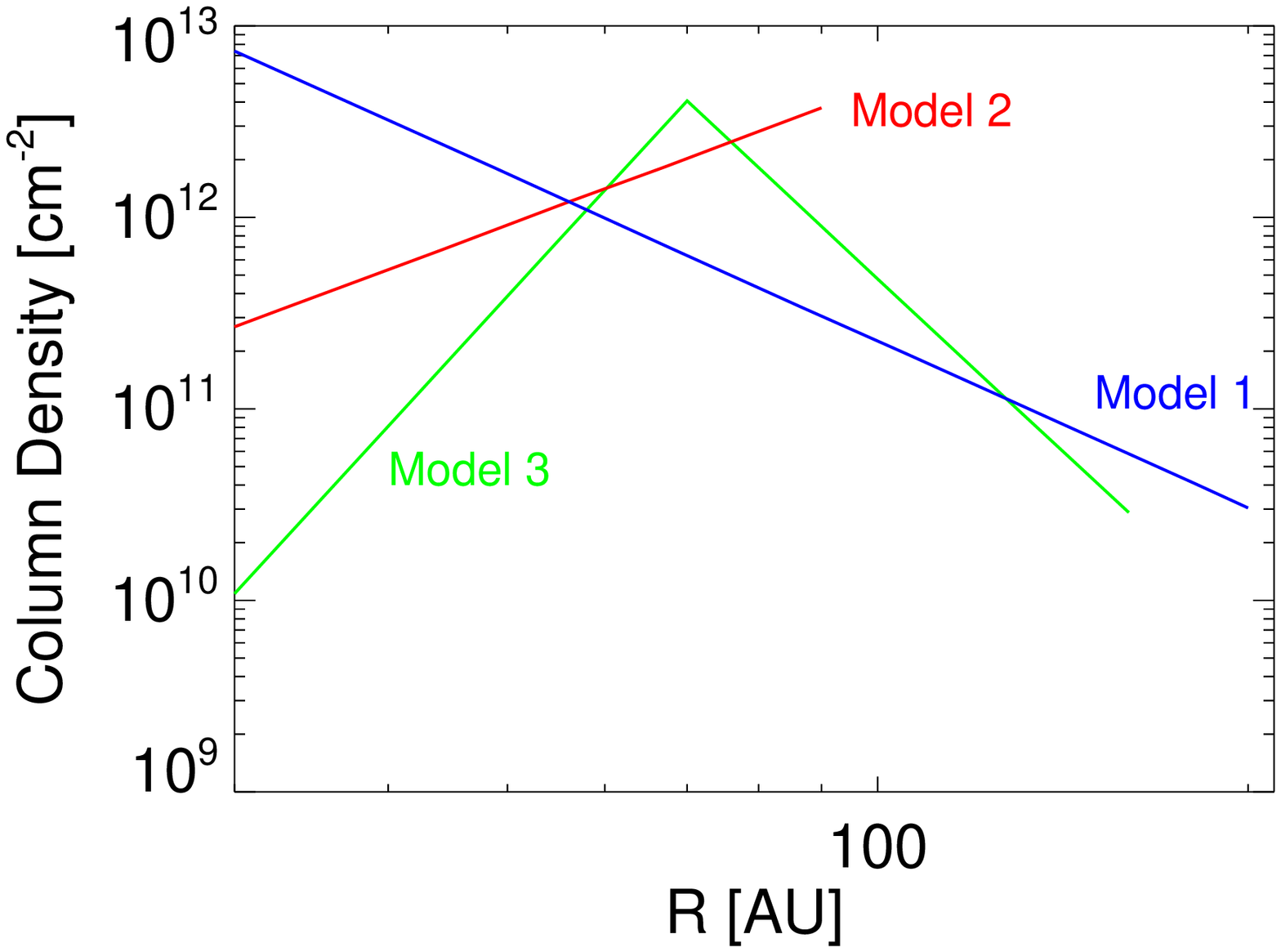}
\caption{Models of DCO$^+$ with different distributions of radial
column densities.
 \label{fig:dcomodels}}
\end{figure}

\begin{figure*}
\centering
\includegraphics[width=0.9\textwidth]{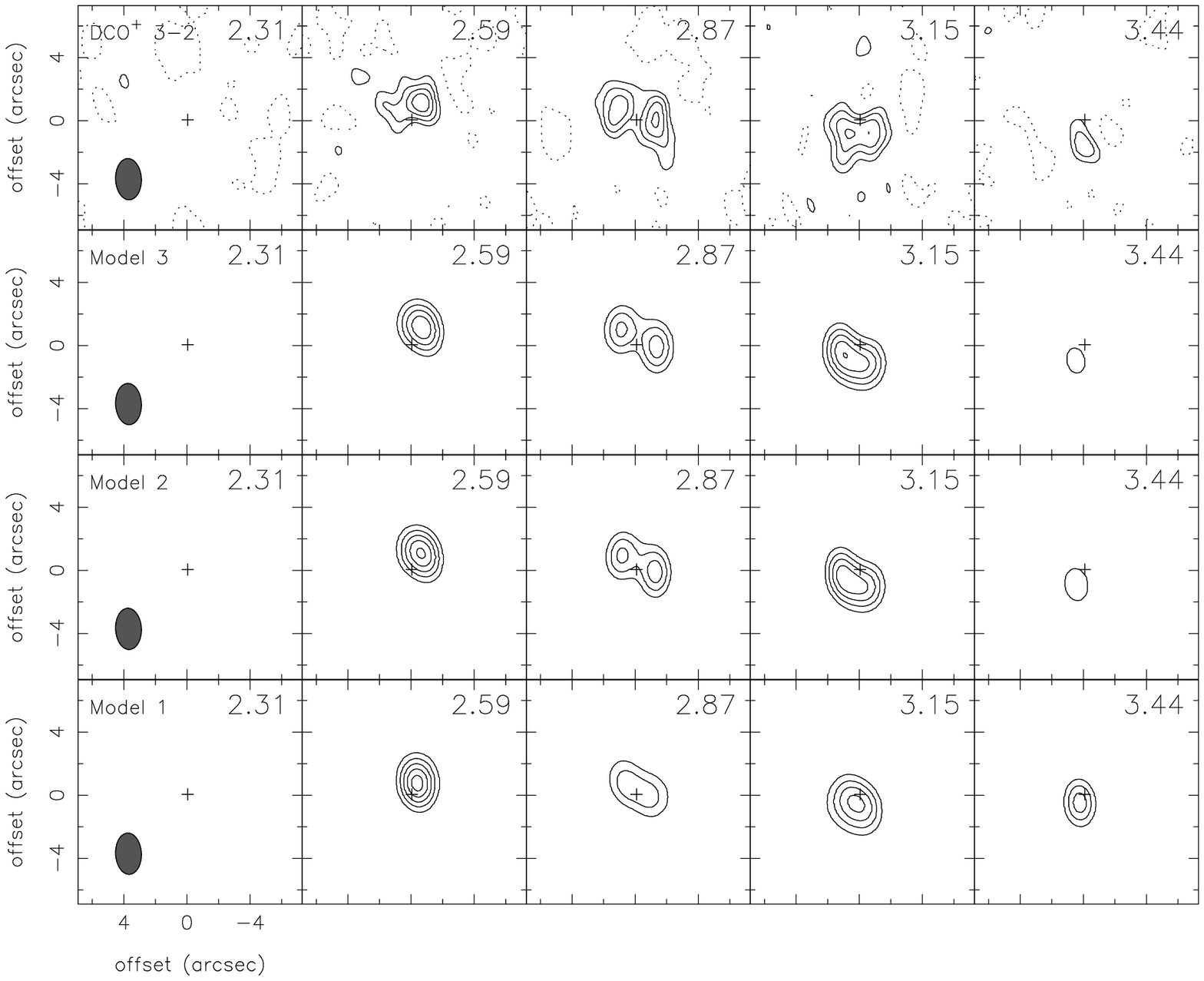}
\caption{DCO$^+$ J=3--2 channel maps toward TW Hydrae and the simulated
model distributions depicted in Figure~\ref{fig:dcomodels}.
 \label{fig:dcomodelmaps}}
\end{figure*}

Examination of the channel maps shows upon inspection that the radial
distributions of DCO$^+$ and HCO$^+$ are different. To demonstrate
how differences in the radial distribution affect the resulting
images, Figure~\ref{fig:dcomodels} presents three models of DCO$^+$ 
radial distribution 
and  Figure~\ref{fig:dcomodelmaps} shows the corresponding simulated 
channel maps of the DCO$^+$ J=3--2 line. 

Model 1 assumes that the DCO$^+$ distribution follows the best-fit model
of HCO$^+$. The minimum $\chi^2$ is determined to be 575396 and
the corresponding DCO$^+$/HCO$^+$ is found to be 4.7 $\times$ 10$^{-2}$.
Comparing the simulated maps from Model 1 with the data
in  Figure~\ref{fig:dcomodelmaps} shows a distinct difference in that the
double peaked nature of the central channel in the data is more obvious
than in this model. The contrast of the contour levels between the central
channel and the adjacent channels are also smaller in the data than in
this model. Because the emission of the central channel mostly
originates at large disk radii,
these differences suggest that the DCO$^+$ emission arising from the
outer regions of the disk is stronger than predicted by this model.
The slope of radial DCO$^+$ distribution does not decrease as steeply 
as does HCO$^+$. In other words, the D/H ratio must increase with
increasing radius. 

Model 2 shows the best-fit result for the radial distribution 
of DCO$^+$ assuming a single power index. As shown in
Figure~\ref{fig:coldens}, the radial distribution of DCO$^+$
is strikingly different from that of HCO$^+$, with a positive power
index of 2.4 and a smaller but better constrained 
R$_{out}$ 90$\pm$5 AU. The contours of the iso-$\chi^2$ surface 
for DCO$^+$ in the middle panel of Figure~\ref{fig:chi} indicate 
that the uncertainty of the radial power index is very large but that
the index is still larger than 1.6 within the 1$\sigma$ error, 
much larger than $-$2.9 found for HCO$^+$.
The simulated DCO$^+$ channel maps for
Model 2 shown in Figure~\ref{fig:dcomodelmaps} are an improvement over
Model 1 in matching the data. However in this model,
R$_{out}$ ($\sim$ 90 AU) is much smaller than the disk radius observed with 
HCO$^+$ and CO ($\sim$ 200 AU) and DCO$^+$ increases
with radius but then disappears sharply at R$_{out}$$\sim$90 AU, as a 
step function, which is hard to understand. Comparison with the data 
shows that there are fewer complete contours around the double peaks 
in the central channel, which suggests that the DCO$^+$ emission 
is maximum at an intermediate radius rather than at the edge. 

\begin{figure*}
\centering
\includegraphics[width=0.9\textwidth]{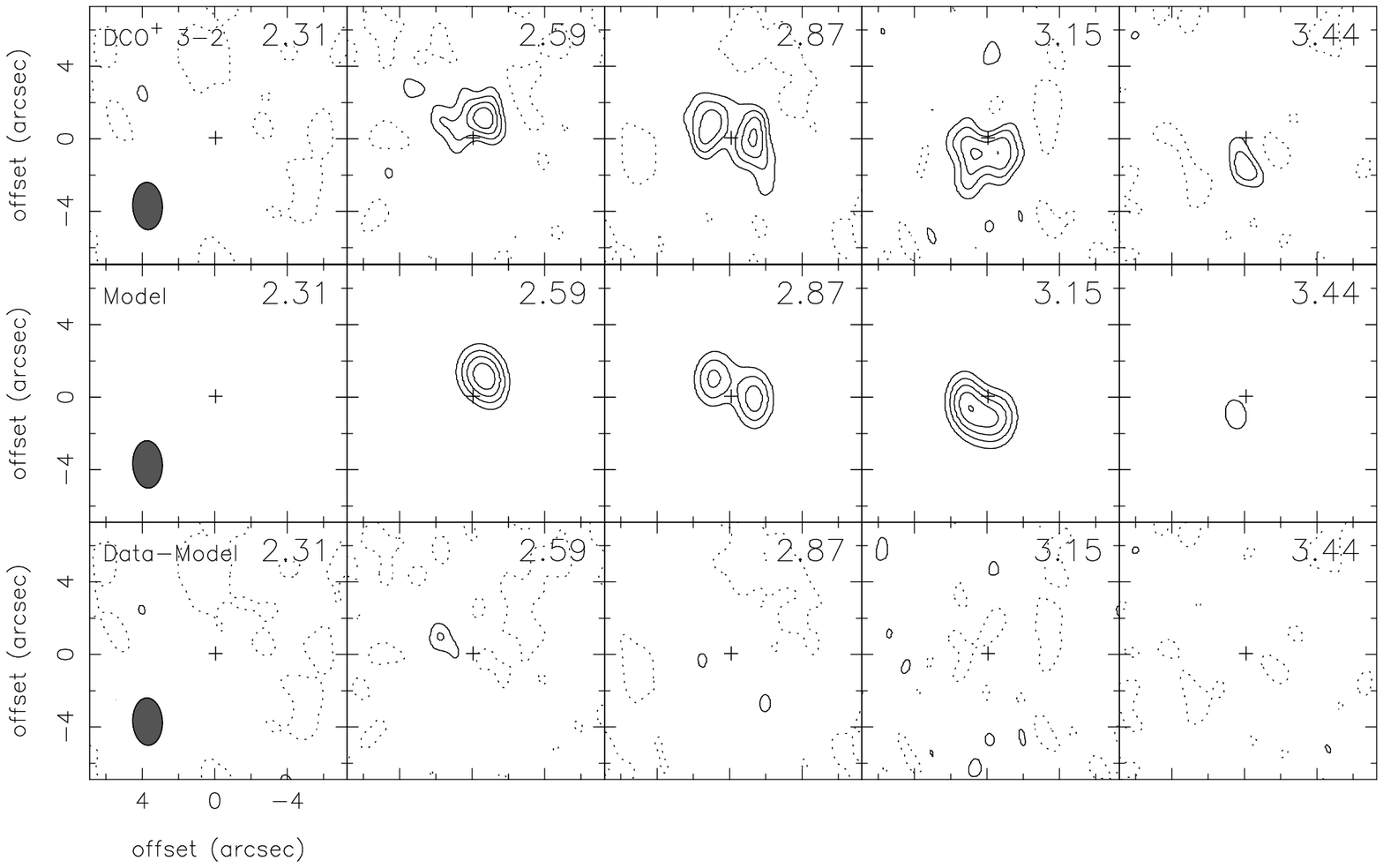}
\caption{Top: Velocity channel map of the DCO$^+$
J=3--2 toward TW~Hydrae. The angular resolution is $2\farcs6 \times
1\farcs6$ at PA  2.8$^{\circ}$. The cross
indicates the continuum (stellar) position. The axes are offsets from
the pointing center in arcseconds.
The 1$\sigma$ contour step is 0.12 Jy beam$^{-1}$ and the
contours start at 2$\sigma$. Middle: channel map of
Model 3 with the same contour levels. Bottom: difference
between Model 3 and data on the same contour scale.
 \label{fig:dcomap}}
\end{figure*}

For Model 3, instead of using a single power index (p) to fit for the radial
distribution of DCO$^+$ column density N(DCO$^+$), the fit uses two power
indices (p1 and p2) and a turning point (R$_t$) where the power law index
changes from p1 to p2, i.e. the location of the peak of N(DCO$^+$). The 
parameters p1, p2 and R$_t$ are searched within limited grids to minimize
$\chi^2$. In this model, N(DCO$^+$) is found to increase with radius out
to $\sim70$~AU, and then to decrease.  Table 3 presents the best-fit 
parameters; no error estimation is provided due to the computation
difficulties. Figure~\ref{fig:dcomodelmaps} shows the best-fit model images. 
Comparison of Model 3 with the data shows a visual improvement over
Model 2 in the central channel, although the $\chi^2$ value of the 
two models are not distinguishable. We thus 
cannot clearly discriminate with the $\chi^2$ 
statistic if there is indeed a peak with radius for N(DCO$^+$) 
as in Model 3,  or if DCO$^+$ increases with radius and disappears 
suddenly around 90 AU as in Model 2. Even with this
ambiguity, however, both models imply that 
the D/H ratios change by at least an order of magnitude (0.01 to 0.1) from
radii $<$30 AU to $>$70 AU and that there is a rapid falloff of 
N(DCO$^+$) at radii larger than 90 AU.
Because the emission in the central channel comes from the 
outer part of the disk (Keplerian rotation) projected along 
the line-of-sight with a very small inclination of around 7 degrees 
for TW~Hydrae, the central velocity channel is most
important for constraining the radial distribution. 
Based on the difference in the central velocity channel map
between the models, we believe Model 3 is the more plausible
description of the radial distribution of DCO$^+$. 
Figure~\ref{fig:dcomap} shows the channel maps of the DCO$^+$ J=3--2 
emission, together with those of Model 3 and the data-model residuals.

Observations of deuterated molecular ions and the level of deuterium
fractionation have been used to estimate the ionization degree in molecular
clouds, and a similar analysis can be applied to circumstellar disks.
If we consider only the
ionization balance determined by HCO$^+$, H$_3^+$, DCO$^+$
and electrons in steady state as shown in Equation 14 of
\citet{caselli02}, the electron fractional abundance can be derived to be 
around 10$^{-7}$. Of course, this value is only valid in the intermediate 
layer where HCO$^+$ is abundant and multiply deuterated H$_3^+$ is
less abundant than HCO$^+$. 
Several important complications are also neglected in this analysis, 
including the presence of other atomic and molecular ions, 
neutral species besides CO which destroy H$_2$D$^+$, and 
negatively charged dust grains and refractory metals. 
Still, accurate measurements of DCO$^+$ and HCO$^+$ are 
the first steps toward an understanding of the ionization 
fraction in the disk.

The increase of D/H ratios from inner to outer disk is generally
consistent with the current theoretical models of the gas-deuterium 
fractionation processes that consider the effect of cold temperature. 
But the quick disappearance of DCO$^+$ at radii beyond 90 AU 
(comparing with R$_{out}$ around 200 AU for CO and HCO$^+$) 
is puzzling, since DCO$^+$ is expected to be abundant and observable 
in the cold outer region of the disk where HCO$^+$ is still available. 
More theoretical work is needed to explain the disappearance of
DCO$^+$ in the outer part of the disk.

\subsection{HCN and DCN}

\begin{figure*}
\centering
\includegraphics[width=0.9\textwidth]{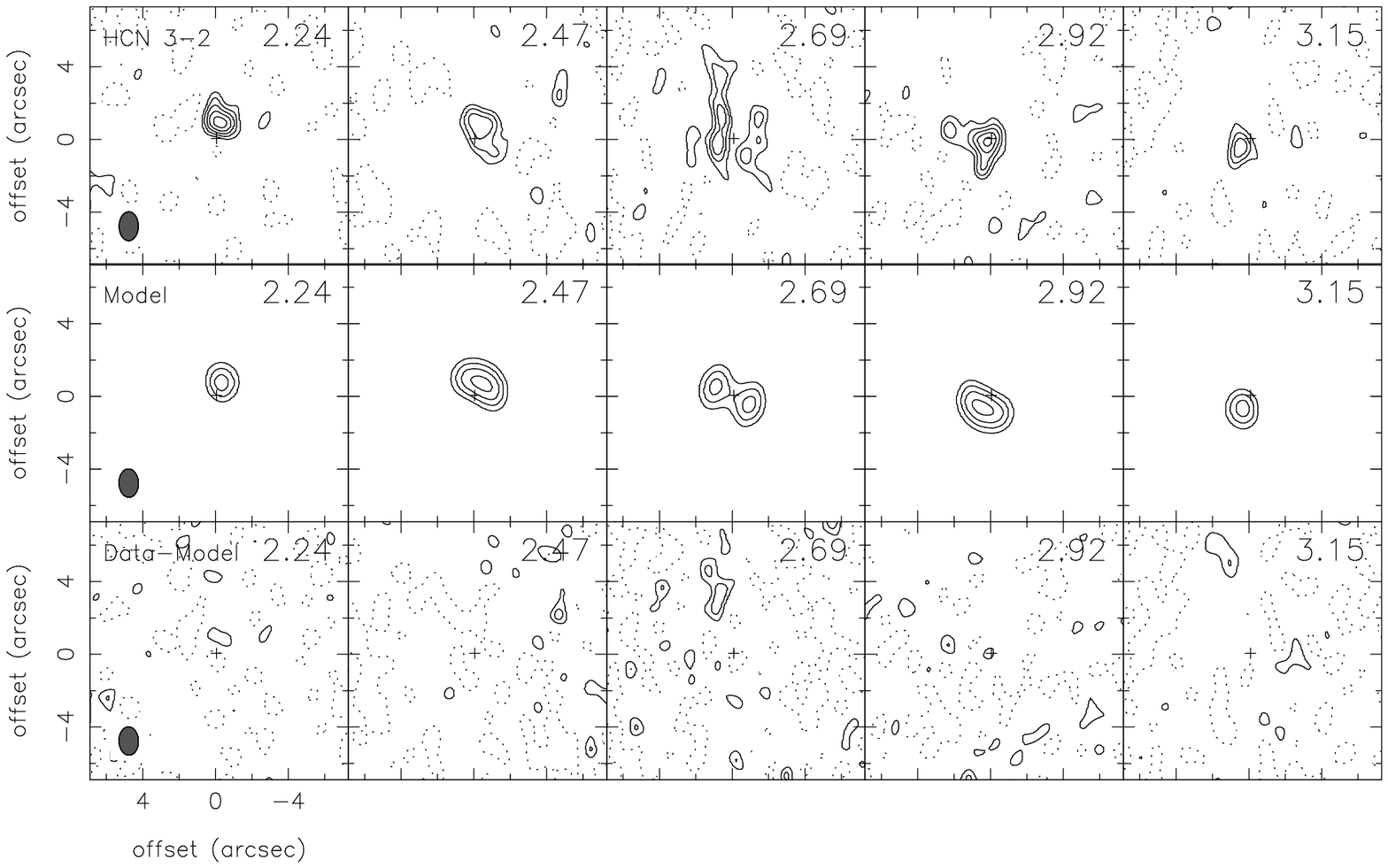}
\caption{Top: Velocity channel map of the HCN
J=3--2 toward TW~Hydrae. The angular resolution is $1\farcs6 \times
1\farcs1$ at PA -0.5$^{\circ}$. The cross
indicates the continuum (stellar) position. The axes are offsets
from the pointing center in arcseconds.
The 1$\sigma$ contour step is 0.35 Jy beam$^{-1}$ and the
contours start at 2$\sigma$. Middle: channel map of the
best-fit model with the same contour levels. Bottom: difference
between the best-fit model and data on the same contour scale.
 \label{fig:hcnmap}}
\end{figure*}

Figure~\ref{fig:hcnmap} shows the HCN J=3--2 channel maps,
together with the best fit model and residuals. Table 3 lists the best-fit 
model parameters, and Figure~\ref{fig:coldens} shows the radial distribution 
of column density derived. The $\chi^2$ surface shown in the bottom
panel of Figure~\ref{fig:chi} indicates R$_{out}$ 100$\pm$10 AU and 
p$_{\rm HCN}$ $-$1.0$\pm$1.2. The radial power index is poorly 
constrained probably due to more complex distributions for HCN. 
A detailed comparison of the molecular distributions of HCN and CN 
will be presented elsewhere.

Although the best fit vertical parameters seem to indicate that
HCN ($\sigma_{m,\rm HCN} = 30$) is found much deeper toward the midplane
than is HCO$^+$ ($\sigma_{m,\rm HCO^+} = 10$), we
emphasize that we are not able to accurately constrain the vertical
distributions from the present data. This ambiguity strongly affects the
column density of HCN (i.e. $\Sigma_{\rm HCN}$(10AU)) needed to fit the
data (not the power index p$_{\rm HCN}$ of radial distribution). A
worse fit to 
the data ($\chi^2$ larger by 3$\sigma$ over the best-fit model) 
can be obtained by assuming that the vertical distributions
of HCN and HCO$^+$ are the same, but the HCN column density is 1.5 times
larger than that needed for the best fit model due to higher density
near the midplane.

\begin{figure*}
\centering
\includegraphics[width=0.8\textwidth]{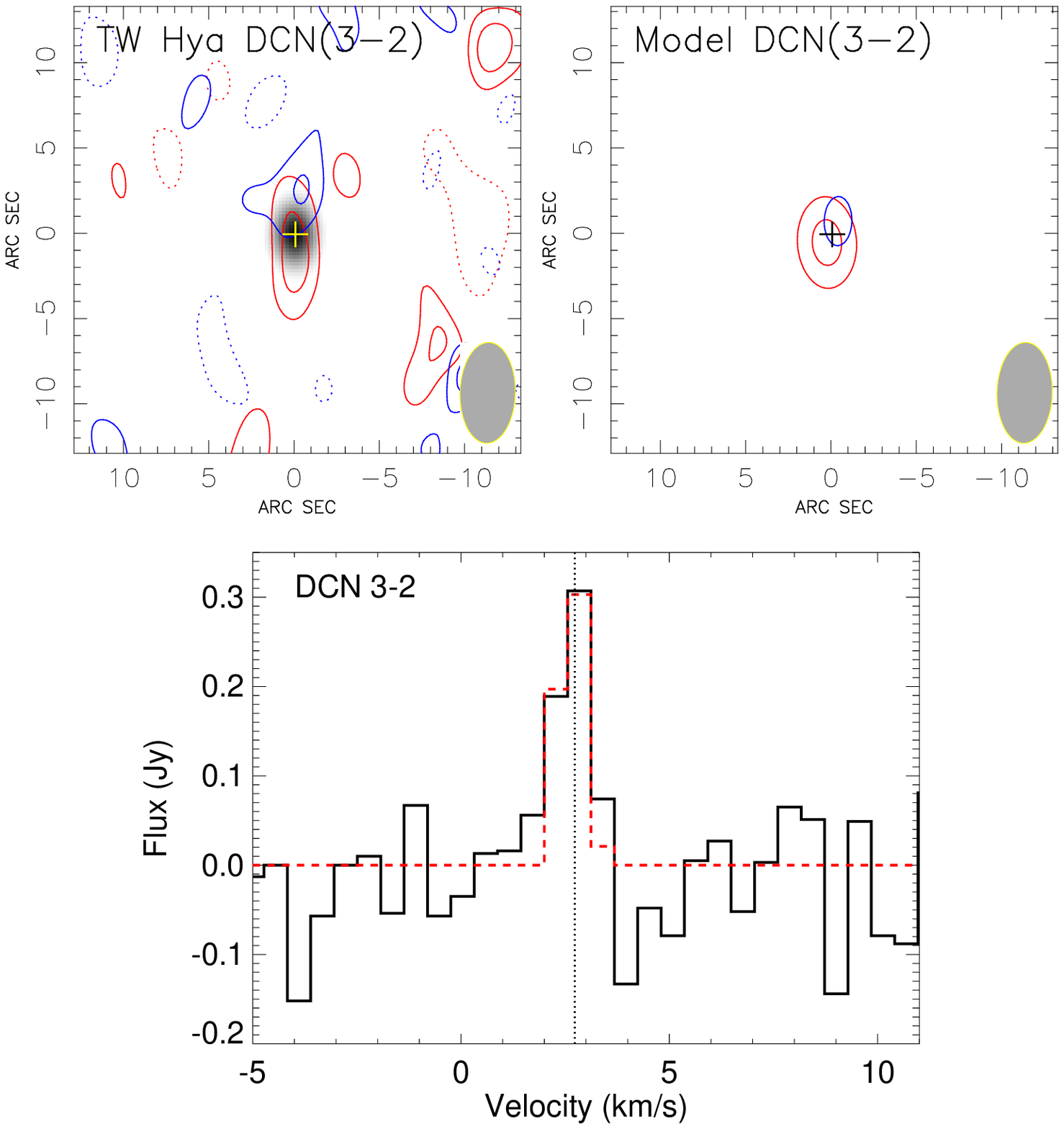}
\caption{Top left: DCN J=3--2  velocity channel maps ({\it red: 2.84 km
s$^{-1}$, blue: 2.28 km s$^{-1}$}) from TW~Hydrae, overlaid on the 217 GHz
dust continuum map ({\it gray scale}). The cross indicates the
position of the continuum peak.
 Top right: the simulated model for DCN
J=3--2. The 1$\sigma$ contour step is 0.09 Jy beam$^{-1}$ and the
contours start with 2$\sigma$. Bottom: the beam-averaged
DCN 3--2 spectra at the continuum (stellar)
position. The SMA data are presented by the solid histogram, and
the simulated model by the dashed histogram. The vertical
dotted line indicates the position of the fitted V$_{LSR}$ for HCN J=3--2.
 \label{fig:dcnmap}}
\end{figure*}

The DCN J=3--2 transition is detected at a signal-to-noise ratio of 3
near the fitted HCN V$_{LSR}$ of 2.73 km~s$^{-1}$ (Figure~\ref{fig:specdata}).
While this signal-to-noise ratio is not high, the significance of the
detection is further supported by the channel maps (Figure~\ref{fig:dcnmap}
upper panel) where the velocity gradient along the disk position angle of
$\sim$ $-$30$^{\circ}$ is consistent with that seen in CO J=2--1 and J=3--2
(\citealp{qi_h04}) and the other molecular lines presented in this paper.
Since the DCN 3--2 emission is weak, we are not able to fit for
the molecular distribution and so make the simplifying assumption that the
distribution of DCN follows that of HCN. As with H$^{13}$CO$^+$, we fit
the DCN/HCN ratio over the whole disk and determine the DCN/HCN ratio to be
1.7$\pm$0.5 $\times$ 10$^{-2}$. To again emphasize the impact of the
assumed vertical distribution on the derived column densities, the
DCN/HCN ratio could be as high as 5 $\times$ 10$^{-2}$ if HCN and DCN are
distributed vertically over the same region as is HCO$^+$.

Highly fractionated DCN/HCN ratios have been measured in comets. In the
coma of comet Hale-Bopp, for example, \citet{meier_o98} reported the
ratio to be around 2.3 $\times$ 10$^{-3}$, but higher DCN/HCN ratios -- 
(D/H)$_{{\rm HCN},jet}$ $\approx$0.025 are detected from the pristine material 
sublimed from icy grains ejected in jets from the nucleus which may present 
a more representative sampling of cometary ices that have not experienced
significant thermal processing (\citealp{blake_q99}).
Such ratios are consistent with those found here in the TW~Hydrae disk,
indicating that high D/H ratios in comets could originate from
material in the outer regions of disks where {\it in situ} deuterium
fractionation is ongoing, rather than requiring an inheritance from 
interstellar material.

\subsection{Upper limits for H$_2$D$^+$, D$_2$H$^+$ and HDO}

In the disk midplane, H$_2$ is expected to be gaseous and the molecular ion
formed by the cosmic ray ionization of H$_2$, H$_3^+$, is expected to
be the most abundant ion. Unfortunately, H$_3^+$ is only detectable in cold
gas via infrared absorption. In its deuterated forms, however, H$_2$D$^+$
and even D$_2$H$^+$ are expected to be abundant in the cold, dense gas
(\citealp{ceccarelli_d05}). 
The ground-state transition of ortho-H$_2$D$^+$ was first detected in a
young stellar object (NGC 1333 IRAS 4A) by ~\citet{stark_v99} and in a 
prestellar core (L1544) by ~\citet{caselli_v03}.
Both H$_2$D$^+$ and D$_2$H$^+$ have been
detected toward another pre-stellar core 16293E via their ground-state submm
rotational lines (\citealp{vastel_p04}). The inclusion of multiply deuterated
H$_3^+$ in chemical models leads to predictions of higher values of the D/H
ratio in cold, high-density regions of the interstellar medium.
Similarly, in the dense, cold disk midplane, CO is depleted, and high
abundances of H$_2$D$^+$ and D$_2$H$^+$ are expected. For this reason,
\citet{ceccarelli_d04} searched for the ground-state transition of ortho-H$_2$D$^+$
with the Caltech Submillimeter Observatory (CSO), and reported 3.2$\sigma$
and 4.7$\sigma$ detections toward TW~Hydrae and DM~Tauri, respectively.
With the 400 GHz and 690 GHz receiver-equipped SMA antennas, we
searched for the 372 GHz ortho-H$_2$D$^+$ 1$_{1,1}$--1$_{1,0}$
and 690 GHz para-D$_2$H$^+$ 1$_{10}$--1$_{01}$ lines toward TW~Hydrae.
No significant emission signals were detected. Here we discuss the
upper limits and their implications.

Our 3-antenna SMA observations give a 1$\sigma$ upper limit for the
ortho-H$_2$D$^+$ 1$_{1,1}$--1$_{1,0}$ line emission of 1.2
Jy~beam$^{-1}$~km~s$^{-1}$  with a $4\farcs7 \times 3\farcs8$ synthesized beam.
To compare the result with single dish data, the extent of the source
emission must be known. Since H$_2$D$^+$
was observed along with the H$^{13}$CO$^+$ 4--3 line in a dual-receiver
observation on 28 December 2006 and H$^{13}$CO$^+$ 4--3 has been clearly
detected at JCMT (\citealp{vandishoeck_t03}), we can compare the
intensities of these two lines between the SMA data and the single
dish observations to constrain the emitting areas. The
H$^{13}$CO$^+$ 4--3 line was detected at JCMT with an integrated
intensity of  0.07 K~km~s$^{-1}$ in a 13$''$ beam
(\citealp{vandishoeck_t03}); while for the SMA the integrated intensity
of this line is determined to be 0.61 Jy~beam$^{-1}$~km~s$^{-1}$ with a
beam of $4\farcs1 \times 1\farcs8$. If the extent of the
emission of H$^{13}$CO$^+$ is similar to that of H$_2$D$^+$, our
interferometric upper limit of Jy~beam$^{-1}$~km~s$^{-1}$ for
H$_2$D$^+$ (considering the small change of the beam sizes) becomes
0.09 K~km~s$^{-1}$ (1$\sigma$) or 0.27 K~km~s$^{-1}$ (3 $\sigma$)
upper limits, which is consistent with 2$\sigma$ upper limits of about
0.2 K~km~s$^{-1}$ by the JCMT (\citealp{thi_v04}) but slightly lower
than the 3.2$\sigma$ detection of 0.39 K~km~s$^{-1}$ by
\citet{ceccarelli_d04}. We estimate the 1$\sigma$ upper limit of the
ortho-H$_2$D$^+$ column density to be 1.7 $\times$ 10$^{12}$ cm$^{-2}$
according to the Equation 4 of \citet{vastel_p04}, which is slightly less 
than the 2$\sigma$ upper limit estimate of 4.4 $\times$ 10$^{12}$ cm$^{-2}$
by \citet{thi_v04} since the SMA data had a smaller noise
level. Of course this analysis assumes the extent of H$_2$D$^+$ is 
similar to that of H$^{13}$CO$^+$. With the deployment 
of more 400 GHz receivers on the SMA and further H$_2$D$^+$ observations, 
it should be possible to provide rather better constraints on the H$_2$D$^+$ 
abundance in the disk.

We estimate the 1$\sigma$ upper limit for para-D$_2$H$^+$ 1$_{10}$--1$_{01}$
to be 5.35 Jy~beam$^{-1}$~km~s$^{-1}$ with a beam of $3\farcs3 \times 1\farcs3$.
The 1$\sigma$ upper limit to the para-D$_2$H$^+$ column density is estimated 
to be  9.0 $\times$ 10$^{14}$ cm$^{-2}$. This is less constrained than 
H$_2$D$^+$ due to the relatively poor system sensitivity at 690 GHz.

For HDO 3$_{1,2}$--2$_{2,1}$, the 1$\sigma$ upper limit is 0.10
Jy~beam$^{-1}$~km~s$^{-1}$. Assuming an excitation temperature of
30 K, the upper limit for the HDO column density is 2.0 $\times$ 10$^{14}$
cm$^{-2}$. Since the lower state energy level of this line is nearly
160 K, it must originate from warm regions of the disk which are
quite distinct from the cold layers where HDO ground state transition
absorption, as found in DM Tauri (\citealp{ceccarelli_hdo}), must arise --
although we note that the detection of the HDO absorption line in DM Tauri 
has been disputed by \citet{guilloteau_p06}.

\section{Summary and Conclusions}

Observations of deuterated species in circumstellar disks are
important to understand the origin of primitive solar system bodies
in that they can directly constrain the deuterium fractionation in
the outer regions where cometary ices are likely formed.
Spatially resolved observations of the D/H ratios in disks enable the
comparison of the fractionation measured in comets such as Hale-Bopp
(\citealp{blake_q99}) with the specific conditions at each disk radius.
We have presented the first images of DCO$^+$ and DCN emission from the
disk around a classical T Tauri star, TW~Hydrae, along with images of
the HCN and HCO$^+$ J=3--2 lines.
The observations of deuterium fractionation serve as a clear measure of
the importance of low-temperature gas-phase deuterium fractionation
processes. These observations strongly support the proposed link among
high gas densities, cold temperature and enhanced deuterium
fractionation. Detailed chemical models are still needed to explain 
how DCO$^+$ disappears from the outer part of the disk. 

The similarity of the D/H ratios in cold clouds, disks and pristine cometary
material has been used to argue that the gas spends most of its lifetime
at low temperatures and is incorporated into the disks before the envelope
is heated, i.e. before the Class I stage. By combining self-consistent
physical models and 2D radiative transfer codes to interpret high spatial
resolution millimeter-wave molecular images, we are only now beginning to
investigate the radial and vertical distributions of molecules in disks.
The radial distribution of DCO$^+$ in the disk of TW~Hydrae indicates
that {\it in situ} deuterium fractionation is ongoing. The molecular
evolution within disks must therefore be considered in the investigation of
the origin of primitive solar system bodies.

We have obtained less stringent constraints on the vertical distributions of
molecules in the disk of TW~Hydrae. To address the ambiguity present in the
analysis of single objects, data from a robust sample of disks is needed,
in particular one that covers a range of disk inclinations. More sensitive
observations are also needed for the rare isotopologues H$^{13}$CN, H$^{13}$CO$^+$ 
and, of course, DCN, to understand the radial and vertical gradient of 
deuterated species in these disks. In the future, observations of DCN 
and other species with the Atacama Large Millimeter Array will provide further 
insight into the chemical state of protoplanetary disks.

\acknowledgements
Partial support for this work comes from 
NASA Origins of Solar Systems Grant NNG05GI81G. 
M.R.H is supported by a VIDI grant from the Netherlands 
Organization for Scientific Research. C.Q. acknowledges Paola Caselli for 
her help and useful suggestions. We thank the referee for very 
useful comments.

\end{document}